\documentstyle[aps,preprint]{revtex}
\begin{document}
\input{psfig}
\draft
\title{\bf Unitary transformation approach for the trapped ion dynamics}
\author{H. Moya-Cessa\footnote{Electronic address: hmmc@inaoep.mx}
\thanks{Permanent address:INAOE, Coordinaci\'on de Optica, Apdo. Postal 51 y 216, 72000 
Puebla, Pue., Mexico}, 
A. Vidiella-Barranco\footnote{Electronic address: vidiella@ifi.unicamp.br}, 
J.A. Roversi\footnote{Electronic address: roversi@ifi.unicamp.br}, 
and S.M. Dutra\footnote{Electronic address: dutra@rulhm1.leidenuniv.nl}}
\address{Instituto de F\'\i sica ``Gleb Wataghin'',
Universidade Estadual de Campinas,
13083-970   Campinas  SP  Brazil}
\date{\today}
\maketitle
\begin{abstract}
We present a way of treating the problem of the interaction of a single trapped ion with
laser beams based on successive aplications of unitary transformations onto the
Hamiltonian. This allows the diagonalization of the Hamiltonian, by means of recursive 
relations, without performing the Lamb-Dicke approximation. 
\end{abstract}

\section{Introduction}

Trapped ions interacting with laser beams \cite{wine0,wiem} have become an extremelly 
interesting system for
the investigation of fundamental physics, e.g. the generation of states of the
harmonic oscillator \cite{cira1,vogel,monr} as well as for potential applications such as 
precision spectroscopy \cite{wine} and quantum computation \cite{cira0}.    
The theoretical treatment of the interaction of a trapped ion with one or 
several laser beams constitutes a complicated problem, being the full Hamiltonian
highly nonlinear. Therefore approximations are normally required, such as, for instance,
the Lamb-Dicke approximation, in which the ion is considered to be confined within a 
region much smaller than the laser wavelength. This makes possible to obtain Hamiltonians 
of the Jaynes-Cummings type, in which the center-of-mass of the trapped ion plays the 
role of the field mode in cavity QED. Recently \cite{ours}, it has been suggested a new 
approach to this problem, based on the application of a unitary transformation which
linearizes the total ion-laser Hamiltonian. Moreover, in that approach \cite{ours} 
it is obtained a 
Jaynes-Cummings type Hamiltonian including counter-rotating terms. A rotating wave
approximation (RWA) may be performed provided we are in the special ``resonant'' regime,
or
$\Omega\approx\nu$, where $\Omega$ is basically related to the laser intensity and $\nu$
is the frequency of the ion in the magneto-optical trap. The RWA performed this way allows 
the exact diagonalization of the transformed Hamiltonian, but on the other hand poses an 
upper limit to the possible values of the Lamb-Dicke parameter. In this paper we propose a
method of diagonalization of the ion-laser Hamiltonian entirely based on unitary
transformations. As a first step we linearize the Hamiltonian, as it is done in 
\cite{ours}. The resulting Hamiltonian is further transformed, and cast in a form that it
is suitable for diagonalization. Although the diagonalization procedure is an approximate
one, based on recursion relations, it is not based in an expansion on any of the 
parameters of the problem. Therefore, the found solution does not depend on any specific 
regime, such as the Lamb-Dicke regime, for instance.

This paper is organized as follows: In section 2 we obtain the linearized form of the 
full Hamiltonian. In section 3 we show how the resulting Hamiltonian may be further 
transformed in order to be diagonalized. In section 4 we summarize our conclusions. 

\section{Linearization of the Hamiltonian}

We consider a single trapped ion interacting with two laser (classical) plane waves 
(frequencies $\omega_1$ and $\omega_2$), in a Raman-type configuration
($\omega_L=\omega_1-\omega_2$). 
The lasers effectively drive the electric-dipole forbidden 
transition $|g\rangle\leftrightarrow|e\rangle$ (frequency $\omega_0$), with
$\delta=\omega_0-\omega_L$. We end up with an effective two-level system for the 
internal degrees of freedom of the atom and the vibrational motion, after 
adiabatically eliminating the third level in the Raman configuration.
This situation is described, in the atomic basis, by the following effective Hamiltonian 
\cite{poya}
\begin{equation}
\hat{H}= \hbar\left(\begin{array}{cc}
	     \nu \hat{n}+\frac{\delta}{2}  &  \Omega e^{i\eta\hat{X}} \\
          \Omega e^{-i\eta\hat{X}} & \nu\hat{n}-\frac{\delta}{2}
		      \end{array}\right),\label{H}
\end{equation}
being $\hat{X}=\hat{a}+\hat{a}^\dagger$, $\hat{n}=\hat{a}^\dagger\hat{a}$,
$\eta$ the Lamb-Dicke parameter, and $\hat{a}^{\dagger}$ 
($\hat{a}$) the ion's vibrational creation (annihilation) operator. 

By applying the unitary transformation
\begin{equation}
\hat{T}=\frac{1}{\sqrt{2}}\left(\begin{array}{cc}
	     \hat{D}^\dagger(\beta)  & -\hat{D}^\dagger(\beta)  \\
          \hat{D}(\beta) & \hat{D}(\beta)
		      \end{array}\right),\label{T}
\end{equation}
to the Hamiltonian in equation (\ref{H}), where $\hat{D}(\beta)=\exp(\beta\hat{a}^\dagger-
\beta^*\hat{a})$ is Glauber's displacement
operator, with $\beta=-i\eta/2$, we obtain the following transformed 
Hamiltonian 
\begin{equation}
\hat{\cal H}
\equiv \hat{T}^{\dagger}\hat{H}\hat{T}  
= \hbar\left(\begin{array}{cc}
	     \nu\hat{n}+\nu {\eta^2 \over 4}+\Omega  & i\lambda\hat{Y}-\frac{\delta}{\eta\nu} \\
          i\lambda\hat{Y}-\frac{\delta}{\eta\nu} & \nu\hat{n}+\nu {\eta^2 \over 4}-\Omega
		      \end{array}\right),\label{TH}
\end{equation}
where $\hat{Y}=\hat{a}-\hat{a}^\dagger$, and $\lambda=\frac{1}{2}{\eta\nu}$.
This result holds for any value of the Lamb-Dicke parameter $\eta$. We have thus 
transformed our original Hamiltonian in equation (\ref{H}) into a Jaynes-Cummings-type 
Hamiltonian, and therefore its exact diagonalization is allowed provided we perform 
the rotating wave approximation (RWA). However this imposes limitations on the 
values of the Lamb-Dicke parameter. Because the ``effective coupling constant'' of
the transformed Hamiltonian is given by $\lambda={1 \over 2}\eta\nu$, we shall have 
${1 \over 2}\eta\nu\ll \nu$ if we want to neglect the counter rotating terms (RWA).
Moreover, we still have to be ``in resonance'', or $\nu\approx 2\Omega$. In this regime
the trapped ion-laser system suitably prepared exhibits long-time-scale revivals 
(superrevivals), as discussed in \cite{ours}. Here we are not going to perform the RWA. 
We propose instead a way of further transforming the Hamiltonian in equation (\ref{TH}) 
in order to allow its diagonalization in an approximate, although nonperturbative way. 
In what follows we are going to present a novel method of transforming the Hamiltonian, 
showing how it may be diagonalized.

\section{Diagonalization of the Hamiltonian}

We have succeeded in transforming the ion-laser Hamiltonian in Equation (\ref{T}) into a 
more tractable form (\ref{TH}). Nevertheless, its exact diagonalization has nor been
achieved yet. The problem is either treated exactly after performing the RWA, or by
means of approximate methods such as perturbative expansions or even numerically 
\cite{sand}. Here we propose a different approach to that problem, based on unitary
transformations. We are going to restrict ourselves to the resonant case $\delta=0$.
After discarding the constant term $\frac{1}{4}\hbar\nu\eta^2$, which just 
represents an overall phase factor, we obtain 

\begin{equation}
\hat{{\cal H}}=\hbar\left(\begin{array}{cc}
	     \nu\hat{n}+\Omega  & i\lambda\hat{Y} \\
          i\lambda\hat{Y} & \nu\hat{n}-\Omega
		      \end{array}\right),\label{THA}
\end{equation}

Now we define the following unitary transformations
\begin{equation}
\hat{T}_1=\frac{1}{\sqrt{2}}\left(\begin{array}{cc}
	     1  & 1  \\
          -1 & 1
		      \end{array}\right),\label{T1}
\end{equation}
and
\begin{equation}
\hat{T}_2=\left(\begin{array}{cc}
	     (-1)^{\hat{n}}  & 0  \\
          0 & 1
		      \end{array}\right).\label{T2}
\end{equation}
These transformations have interesting effects when applied to the Jaynes-Cummings type 
Hamiltonian in (\ref{THA}). First we apply $\hat{T}_1$, which yields
\begin{equation}
\frac{1}{2}\left(\begin{array}{cc}
	     1  & -1  \\
          1 & 1
		      \end{array}\right)\left(\begin{array}{cc}
	     \Omega  & \lambda\hat{Y}  \\
          \lambda\hat{Y} & -\Omega
		      \end{array}\right)\left(\begin{array}{cc}
	     1  & 1  \\
          -1 & 1
		      \end{array}\right)=\left(\begin{array}{cc}
	    -i\lambda\hat{Y}  & \Omega \\
                  \Omega  & i\lambda\hat{Y} 
		      \end{array}\right),
\end{equation}
i.e., the operators are transposed to the diagonal. Then we apply $\hat{T}_2$ to the 
matrix which resulted form the operation above, such that
\begin{equation}
\left(\begin{array}{cc}
	     (-1)^{\hat{n}}  & 0  \\
          0 & 1
		      \end{array}\right)\left(\begin{array}{cc}
	    -i\lambda\hat{Y}  & \Omega \\
                  \Omega  & i\lambda\hat{Y} 
		      \end{array}\right)\left(\begin{array}{cc}
	     (-1)^{\hat{n}}  & 0  \\
          0 & 1
		      \end{array}\right)=\left(\begin{array}{cc}
	    i\lambda\hat{Y}  & \Omega(-1)^{\hat{n}} \\
                  \Omega(-1)^{\hat{n}}  & i\lambda\hat{Y} 
		      \end{array}\right),
\end{equation}
where it has been used the fact that $(-1)^{\hat{n}}\hat{a}(-1)^{\hat{n}}=-\hat{a}$ 
(the same for $\hat{a}^\dagger$). We have now that $\hat{Y}$ is multiplied by the 
identity
matrix. At this stage we apply $\hat{T}_1$ again, rearranging the terms in such a way that
we obtain a Hamiltonian diagonal in the atomic state basis, or
\begin{equation}
\left(\begin{array}{cc}
	     1  & -1  \\
          1 & 1
		      \end{array}\right)\left(\begin{array}{cc}
	    i\lambda\hat{Y}   & \Omega(-1)^{\hat{n}}  \\
          \Omega(-1)^{\hat{n}} & i\lambda\hat{Y}
		      \end{array}\right)\left(\begin{array}{cc}
	     1  & 1  \\
          -1 & 1
		      \end{array}\right)=i\lambda\hat{Y}+
		      \left(\begin{array}{cc}
	  -\Omega(-1)^{\hat{n}}   & 0 \\
                  0  & \Omega(-1)^{\hat{n}} 
		      \end{array}\right).
\end{equation}
Diagonalization of the total Hamiltonian becomes an easy task now, 
because the atomic part is already in a diagonal form. The total transformed Hamiltonian 
may be conveniently expressed in an operator form as
\begin{equation}
\hat{{\cal H}}'=\hat{T}^\dagger_1 \hat{T}^\dagger_2 \hat{T}^\dagger_1 \hat{T}^\dagger 
\hat{H}\hat{T}\hat{T}_1\hat{T}_2\hat{T}_1\label{TTH}=
\hbar\left(\nu\hat{n} + i\lambda\hat{Y}-
\Omega\sigma_z(-1)^{\hat{n}}\right).\label{TF}
\end{equation}
A general expression for the eigenstates of the
Hamiltonian in expression (\ref{TF}) is
\begin{equation}
|\Psi_l\rangle=|\varphi^g_l\rangle|g\rangle +|\varphi^e_l\rangle|e\rangle,
\label{EM}
\end{equation}
with $|\varphi^e_l\rangle=\sum_n C_{n,l}^e |n\rangle$ and 
$|\varphi^g\rangle=\sum_n C_{n,l}^g|n\rangle$. 
From the eigenvalues equation $\hat{{\cal H}}'|\Psi_l\rangle=\Lambda_l |\Psi_l\rangle$ we 
obtain
\begin{equation}
\left(\nu\hat{n}+i\lambda\hat{Y}-\Omega(-1)^{\hat{n}}\right)\sum_n C_{n,l}^e|n\rangle
=\Lambda_l\sum_n C_{n,l}^e|n\rangle,
\end{equation}
and
\begin{equation}
\left(\nu\hat{n}+i\lambda\hat{Y}+\Omega(-1)^{\hat{n}}\right)\sum_n C_{n,l}^g|n\rangle=
\Lambda_l\sum_n C_{n,l}^g|n\rangle.
\end{equation}
The expansion coefficients $(C's)$ may be obtained by means of recursion relations. 
For instance, for the coefficient $C_{l,n}^e$. From equation (\ref{EM}) we may write
\begin{equation}
\sum_n \left( nC_{n,l}^e |n\rangle +i\lambda C_{n,l}^e\sqrt{n} |n-1\rangle -i\lambda C_{n,l}^e
\sqrt{n+1} |n+1\rangle 
- \Omega (-1)^n C_{n,l}^e |n\rangle \right)=\Lambda_l\sum_n C_{n,l}^e |n\rangle.\label{sum}
\end{equation}
After rearranging some of the terms, we have the following relation between the
coefficients
\begin{equation}
C_{n+2,l}^e = \frac{\sqrt{n+1}}{\sqrt{n+2}} C_{n,l}^e -\frac{i}{\lambda}
\frac{\left[\Lambda+\Omega (-1)^{n+1}-\nu(n+1)\right]}{\sqrt{n+2}} C_{n+1,l}^e.
\end{equation}
By multiplying expression (\ref{sum}) by $\langle 0|$, we also obtain the first term
\begin{equation}
C_{1,l}^e=-i\frac{\Omega+\Lambda}{\lambda}C_{0,l}^e.
\end{equation}
Similar relations may be found for the coefficients $C_{n,l}^g$. Of course 
the normalization condition $\sum_n |C_{n,l}^g|^2+\sum_n |C_{n,l}^e|^2=1$ should be satisfied.
Having diagonalized the transformed Hamiltonian, the evolution of the state vector 
becomes a trivial task. For that we have to express a generic state $|\psi\rangle$ in terms of 
the basis states, or
\begin{equation}
|\psi\rangle= \sum_l A_l |\Psi_l\rangle.
\end{equation}
The choice of a specific initial state determines the set of
coefficients $A_l$. 
For instance, if we initially prepare the trapped ion in the state
$|\Psi(0)\rangle=\frac{1}{\sqrt{2}}|\beta\rangle\left(|g\rangle-|e\rangle\right)$, the 
transformed state will read
\begin{equation}
|\psi(0)\rangle=\hat{T}^\dagger_1 \hat{T}^\dagger_2 \hat{T}^\dagger_1 \hat{T}^\dagger
|\Psi(0)\rangle=|0\rangle|e\rangle=\sum_l A_l |\Psi_l\rangle. \label{coef}
\end{equation}
The coefficients $A_l$ may be determined by multiplying the left hand side of Equation 
(\ref{coef}) by $\langle n|\langle e|$ and $\langle n|\langle g|$, so that we obtain a
set of coupled equations for the coefficients $A_l$. It is now easy to calculate the 
time-evolution of the state vector, namely
\begin{equation}
|\psi(t)\rangle= \exp(-i\hat{{\cal H}}'t/\hbar) \sum_l A_l |\Psi_l\rangle=
\sum_l A_l \exp(-i\Lambda_l t/\hbar)|\Psi_l\rangle.
\end{equation}   
The next step is to apply on $|\psi(t)\rangle$ the sequence of transformations 
$\hat{T}^\dagger_1 \hat{T}^\dagger_2 \hat{T}^\dagger_1 \hat{T}^\dagger$ backwards in 
order to recover the wanted solution for the state vector which describes
the trapped ion system interacting with the laser beams.

\section{Conclusions}

We have presented a novel way of treating the problem of the interaction of trapped
ions with laser beams entirely based on unitary transformations. The system
Hamiltonian is successively modified and cast in a more tractable form. We propose a 
method of diagonalization of the transformed Hamiltonian by means of the construction of 
recursive relations for expansion coefficients in the Fock state basis. Despite of 
not being an exact diagonalization, there is no need of performing any approximation, such as
the Lamb-Dicke limit, for instance. 
Interesting possibilities for the investigation of such a system in
different regimes are opened up, given that there are no restrictions on the relevant
parameters and no approximations have been made so far. 

\acknowledgements

This work was partially supported by Consejo Nacional de
Ciencia y Tecnolog\'\i a (CONACyT), M\'exico, Conselho Nacional de 
Desenvolvimento Cient\'\i fico e Tecnol\'ogico (CNPq).

\end{document}